\documentclass{article}
\usepackage{ijcai17}
\usepackage{amsmath}
\usepackage{times}
\usepackage{graphicx}

\title{Personalized Music Recommendation with Triplet Network}
\author{Donghuo Zeng \textsuperscript{1}, Haoting Liang\textsuperscript{2}, Yi Yu\textsuperscript{3}, Keizo Oyama\textsuperscript{4}\\ 
National Institute of Informatics, SOKENDAI, Tokyo, Japan. \\
\textsuperscript{1}zengdonghuo@nii.ac.jp, \textsuperscript{2}s8halian@stud.uni-saarland.de, \textsuperscript{3}yiyu@nii.ac.jp,\textsuperscript{4}oyama@nii.ac.jp}

\begin{document}

\maketitle

\begin{abstract}
Since many online music services emerged in recent years so that effective music recommendation systems are desirable. Some common problems in recommendation system like feature representations, distance measure and cold start problems are also challenges for music recommendation. In this paper, I proposed a triplet neural network, exploiting both positive and negative samples to learn the representation and distance measure between users and items, to solve the recommendation task.    

\end{abstract}

\section{Introduction}
As summarized by \cite{CampbelMusic},common recommendation algorithms, like collaborative filtering-based recommendation and content-base recommendation, only user history information or items features are considered. These kind of algorithms relies on users history which usually lead to sparsity matrix problem. In our solution, we try to use cross-modal information\cite{AudioVisual; yu2018category; yu2017deep}, training data on both user preference and item features at the same time to learn their effective representation for this problem and relationship directly. What's more, in most recommendation systems, negative feedback,which means users dislike that items, are ignore. In this work, we use not only positive feedback but also negative feedback together to embrace more information of the users preference.

Inspired by \cite{DBLP:journals/corr/LeiLLZL16} , we study a three branches network for the music recommendation task. One subnetwork is for user preference and two sub-networks with shared parameters are for positive items and negative items respectively. All three subnetworks will map user preference and items to the same latent semantic space. The final objective is that the distance between user preference and positive items should be closer than that between user preference and negative items. By optimizing the network according to this objective, we want to learn the mapping subnetwork for both users and items, as well as the distance measure function between user preference and items. 

\section{Problem Formulation}
The main idea of the triplet network is to map both music features and user tags to a common feature space, therefore the input of the network is a triplet:user preference, positive item and negative items. Assuming there are n user samples, then the input data should be
\begin{equation*}
(U_{t},I_{t}^{+},I_{t}^{-}),t=1,...n
\end{equation*}
The network exploits both positive and negative items with the objective that in the common space
\begin{equation*}
\mathcal{D}(\pi(\text{User}),\phi(\text{positive items})) < \mathcal{D}(\pi(\text{User}),\phi(\text{negative items}))
\end{equation*}
where $\pi()$ is the mapping function for users and $\phi()$ is for items. $\mathcal{D}()$ is a distance function to measure the distance between users and items in the latent common space. To fulfill this learning objective, we can perceive this objective as a binary classification problem. Let
\begin{equation*}
o_{ij}^{U_t} = \mathcal{D}(\pi(U_t),\phi(i)) - \mathcal{D}(\pi(U_t),\phi(j))
\end{equation*}
where i and j stand for items, and then apply sigmoid function on $o_{ij}^{U_t}$
\begin{equation*}
P_{ij}^{U_t} = sigmoid(o_{ij}^{U_t})
\end{equation*}
which means if item i is closer to the user than item j, where $o_{ij}^{U_t}$ is negative, $P_{ij}^{U_t}$ will drop to 0, otherwise it will grow to 1.

We can consider $P_{ij}^{U_t}$ as labels of a binary classification problem. If item i is the positive item and j is the negative item, the label is 0, otherwise the label is 1. In formal definition, the true label should be
\begin{equation*}
\overline{P}_{ij}^{U_t} = \begin{cases}
0& (i=I_t^{+},j=I_t^{-})\\
1& (i=I_t^{-},j=I_t^{+})
\end{cases}
\end{equation*}
Finally, the problem become a binary classification problem: The input is a paired items(positive and negative) and user preference information. The goal is to learn the mapping network for positive, negative and user tags which can classify whether the paired data is a pos-neg or a neg-pos pair.

Correspond to this formulation, the binary cross-entropy loss function can be used
\begin{equation*}
min_{\pi,\phi,\mathcal{D}}\mathcal{L}({\mathcal{N}}) = \sum_{t}-\overline{P}_{ij}^{U_t}log(P_{ij}^{U_t})-(1-\overline{P}_{ij}^{U_t})log(1-P_{ij}^{U_t})
\end{equation*}

\section{Network Architecture}
\begin{figure}
\includegraphics[width=8.7cm, height=3cm]{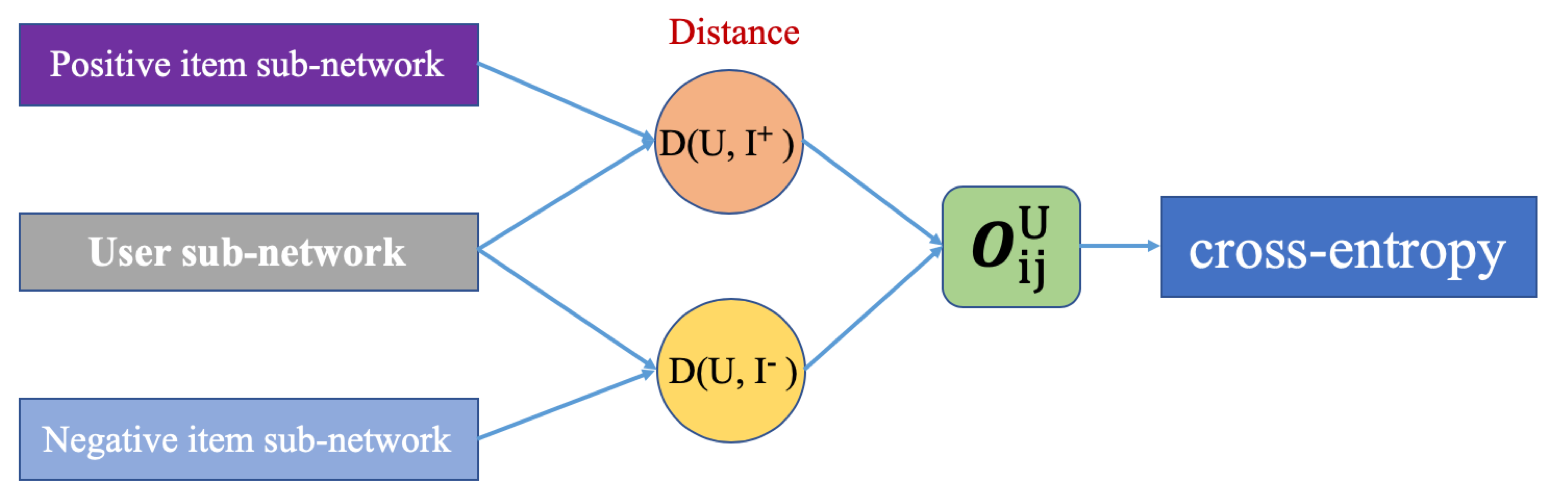}
\centering
\caption{Simplified Network Architecture}
\label{fig:network}
\end{figure}
The triplet neural network is of three sub-networks. Fig. \ref{fig:network} shows a simplified version of the network. One branch is for user tags and two branches are for audios, the upper one for positive audios and the bottom one for negative audios. The two branch share the same parameters. About the distance function $\mathcal{D}$, the element-wise difference of user and item vectors in the latent common space are calculated first and followed by element-wise square operation. Then the squared difference vector is fed into a fully connected layer to get the final weighted distance. What we want to learn from the training data, is the weights for these three subnetworks and the fully connected layer for calculating distance.

\begin{table}[]
    \centering
    \setlength{\tabcolsep}{10mm}
    \begin{tabular}{c|c}
    \hline
    Methods& Accuracy \\
    \hline
Triplet& $57.53\%$\\
\hline
Twonet& $48.24\%$\\
\hline
    \end{tabular}
    \caption{Comparison between different methods}
    \label{tab:a}
\end{table}
When making decision for recommendation for a new user, only one of the subnetwork for items and the subnetwork for user preference are used. After mapping the new user to the latent common space, the distance between the new user to existing items will be calculated, then the nearest items will be returned. 
\begin{table}[]
    \centering
    \setlength{\tabcolsep}{10mm}
    \begin{tabular}{c|c}\hline
        dataset & Accuracy \\
        \hline
        Unbalanced & $57.53\%$ \\
        \hline
        Balanced & $60.04\%$\\
        \hline
        1-to-n balanced & $62.89\%$\\ \hline
    \end{tabular}
    \caption{Different Datasets}
    \label{tab:b}
\end{table}

\begin{table}[]
    \centering
    \setlength{\tabcolsep}{10mm}
    \begin{tabular}{c|c}
    \hline
    Methods& Accuracy \\
    \hline
Triplet& $87.42\%$\\
\hline
Twonet& $71.89\%$\\
\hline
    \end{tabular}
    \caption{Comparison between different methods}
    \label{tab:c}
\end{table}
\section{Dataset}
The dataset consists of more than 26000 songs with their social tags crawled from last.fm. Since the social tags are tagged by users, we apply LDA topic model over the social tags to get 7 top topics and represent users with these topics, which means a users preference is represented by a 10-dim vector here. The 30 seconds songs are processed by mfcc, and finally we get 20 frames for one song and 378-dims for one frame.     
The two sub-networks for music are fully connected networks with 4 hidden layers, and item features are flatten to 7560-dim vector before being fed into the network. Because of the user vectors is 7-dims, the dimensionality of the common space is 7, so the output is 7-dims. The sub-network for user tags is also fully connected network, with 4 hidden layers and the output is also 7-dims.

\subsection{Experiment}
During training stage, pairs of data, pos-neg or neg-pos are fed into the network with labels 0 or 1 to learn the mapping and distance function. 
The first experiment is to test that given a vector of user tags, the system should retrieve the nearest audios, and evaluate its performance by calculating precision, which means among all the test samples, how many returned audios have the same tags with that of the user's interest. Its performance is compared with two branches neural network, which use only the positive items information, using the binary loss to evaluate the performance. The subnetworks of the two branches NN is the same structure as the triplets NN except for it use only two subnetworks for training. Normalization and Dropout are applied to the output of every layer, probabilities of dropout are set to 0.2. The network are trained 200 epochs with batch size 256.  
Table.\ref{tab:a} shows the result of these two methods. The performance of Triplet is better than two branches network.

The second experiment is to test the network on dataset with different distributions. "Unbalance" means all tags combinations have the same number of data pairs. "1-to-n" means we choose multiple negative samples for one positive samples. In previous dataset, each positive item is only match one negative item and in this dataset I choose 10 negative items for each positive items. Table.\ref{tab:b} shows the result on different datasets. It shows 1-to-n balanced dataset have the best performance and 1-to-n dataset the worst. 

The third experiment is that given an audio, the mapping sub-network for audio will map audios to the space and then retrieve nearest audios. Table.\ref{tab:c} shows the result of these two methods. Both given user preference to retrieve audios and given audios to retrieve audios, the triplet network shows better performance than the two branches network.

\section{Conclusion}
From the result, we can come to conclusion that by exploiting both positive and negative items to train the network, a good mapping and distance measure functions can be learn. In the future, more structures of the subnetwork and more types of distance measure can be investigated to seek better performance.

\bibliographystyle{named}
\bibliography{ijcai17}

\end{document}